\documentclass[showpacs,11pt]{revtex4}  
\usepackage{amsfonts}
\usepackage{amssymb,epsf}
\usepackage{latexsym}

\newcommand{\bear}{\begin{eqnarray}}
\newcommand{\eear}{\end{eqnarray}}
\newcommand{\be}{\begin{equation}}
\newcommand{\ee}{\end{equation}}
\newcommand{\bearst}{\begin{eqnarray*}}
\newcommand{\eearst}{\end{eqnarray*}}

\begin{document}

\title{The mass and the coupling of the Dark Particle} 

\author{Elcio Abdalla}
\email{eabdalla@fma.if.usp.br} \affiliation{Instituto de Fisica,
Universidade de Sao Paulo, C.P.66.318, CEP 05315-970, Sao Paulo}

\author{Bin Wang}
\email{wangb@fudan.edu.cn} \affiliation{Department of Physics,
Fudan University, 200433 Shanghai}

\begin{abstract}
We argue that Dark Matter can be described by an interacting field theory
with a mass parameter of the order of the proton mass and an interaction
coupling of the order of the QED coupling.
\end{abstract}


\maketitle

\section{Introduction}

We know that Dark Matter (DM) clumps around galaxies, tracing the
clumping of matter. However, it does not seem to take part in stars and
might not show up in our vicinity, at least not extensively. Can we 
get constraints from these facts? Are they consistent with a 
non interacting Dark Matter (DM) and Dark Energy (DE) model? 
We argue that we need a field theory in
interaction in order to be able to explain such a phenomenology. Moreover,
we believe that the strenght of the interaction as well as the mass of the
Dark Particle are restricted from general arguments.

Non interacting DM can be hardly understood in terms of
field theories. Moreover, matter clumping depends crucially on
friction: a frictionless matter cannot clump, it would just scatter as a
consequence of gravity, while matter with dissipation falls into energy 
traps leading to structures. We thus need an interacting field theory 
describing DM. The fundamental question is how to entail interaction 
from the knowledge of astrophysical phenomenology. 

Some works are quite definite on the requirement of DE and DM
interaction. Several papers dealt with the matter of interaction in 
general  \cite{inter} as well as \cite{interholo} within the holographic
model \cite{holo}. Recently it has been argued \cite{farrarrosen} that
interactions should be present to explain the Bullet observations
\cite{bullet}. Here we give further arguments to support the picture of
interacting DM, possibly with DE. 

\section{Galaxy Formation}

Let us suppose that Dark Matter scatters through some mechanism
described by a cross section $\sigma_{DM}$. In the case there is a Quantum 
Field Theory behind the scene, the fine structure
constant corresponding to the coupling of DM, given by $\alpha_{DM}$, 
should be related to the coupling $g_{DM}$
of DM to some mediator via $\alpha_{DM}=g_{DM}^2$, up to a numerical
factor of order of unity. We also
suppose that the DM field is characterized by two mass
parameters $m_1$ and $m_2<<m_1$ describing fields in interaction. The
motivation is to describe a dark sector as similar as possible to our {\it
  bright} (or baryonic plus leptonic) matter.

Similarly to Thompson scattering, we assume that there is a
relation between the above quantities given by
\bearst
\sigma_{DM} \approx \frac{\alpha_{DM}^2}{m_2^2 }\quad .
\eearst

In such a case, supposing the cooling of DM in very large scales
to be due to some kind of bremsstrahlung process, the cooling time
is \cite{padmanabhan}
\bearst
t_{cool}\approx (n\alpha_{DM}\sigma_{DM})^{-1}\left( \frac Tm_2\right)^{1/2}
= \frac{m_2^2}{\alpha^3_{DM}n}\left( \frac Tm_2\right)^{1/2}\quad ,
\eearst
where $n$ is the DM number density.
This should be compared to the time of gravitational collapse,
\bearst
t_{grav}=\left( \frac{R^3}{GM} \right)^{1/2}\quad ,
\eearst
where now $M$ is the total mass and $R$ the corresponding size. The
temperature is $T = \frac {GMm_1}R$; equating $t_{cool}$ and $t_{grav}$ as a
condition for triggering structure formation and using $n=\frac N {4/3\pi
  R^3}$ and $N=M/m_1$, we have 
\bearst 
R_{dark\; gal}\approx \frac 14\frac{\alpha_{DM}^3}{ G (m_1m_2)^{3/2}} \quad ,
\eearst
We can compare 
with the similar result of baryonic galaxy formation \cite{padmanabhan}
(actually just replacing $m_{1,2}$ by $m_{p,e}$ and $\alpha_{DM}$ by the
fine tructure constant  $\alpha$. For  $R\approx R_{gal}$ we have
($m_e=10^{-3}m_p$) 
\be 
\alpha_{DM}\approx 30 \frac {\sqrt{m_1m_2}}{m_p}\alpha \quad .
\ee
Let us now examine the galaxy mass. Using
$M_{dark\; gal}\approx Nm_1$ and the temperature as
$T \approx  \alpha^2_{DM}m \approx \frac{GMm}{R}$ we find, for $M_{dark\;
  gal}\approx 10 M_{bright\;  gal}$
\be
 \alpha_{DM}\approx 3\alpha\; , \qquad  {\rm consequently
,}\quad 
\sqrt{m_1m_2}\approx 0.1 m_p\quad .
\ee
Therefore the Dark Matter coupling should difer from the QED fine
structure constant by a factor of 3, approximately.

\section{Star Formation}

Supposing that the dark paticles are confined within a small region $d$, their
kinetic energy is of the order $\frac{p^2}{m_2}\approx \frac
1{m_2d^2}$. Their energy is $U\approx \frac{GM^2}R$, their thermal energy
$K\approx NT$ thus $T\approx \frac{GMm_1}R$. The typical distance is
gotten from $d\approx RN^{-1/3}$, thus $T\approx N^{2/3}Gm^2/d$.
Arguing that we have a balance between thermal and kinetic energy
\cite{padmanabhan}, we find $T\approx N^{2/3}Gm_1^2/d-1/(m_2d^2)$, whose
maximum is achieved when 
\bearst 
d&=&d_{dark\; star} =2N^{-2/3}\frac 1{Gm_1^2m_2} 
\eearst
The corresponding temperature turns out to be
\bearst 
T_{dark\; star} \approx N^{4/3}G^2m_1^4m_2 \quad . 
\eearst

We wish to impose that the star ignites.
This is actually a very delicate point. Does it
ignite? It would  emit which kind of energy? In case we have a DE and DM
interaction of the kind suggested elsewhere \cite{inter,interholo} the
decay of DM into DE in average is not preferred (the interaction seems to
be preferable in the other direction for entropic reasons). In case
we first suppose that the physical process is the same as in the baryonic
case, we need an energy $\epsilon \approx \alpha_{DM}^2 m_1$ to ignite the
star, in which case the condition becomes, for the number of particles,
\be 
N_{dark\; star}\approx \left( \frac{\alpha_{DM}}{G} \right)^{3/2}\frac
1{m_1^{9/4}m_2^{3/4}} 
\ee
Comparing with a baryonic (bright) star, $N_{dark\; star}\approx \left(
  \frac{m_p}{m_1} \right)^{3/2} N_{bright\;star} $
The corresponding mass is of the order
\be 
M_{dark\; star}\approx   \left( \frac{m_p}{m_1} \right)^{1/2}M_
{bright\; star} 
\ee
We also compute the radius of such a star,
\bearst 
R_{dark\; star} \approx N_{dark\; star}^{-1/3}\frac 1{Gm_1^2m_2} \quad .
\eearst
For the range of values $m_1\approx 3m_p$ and $m_2\approx 3m_e$,
compatible with the previous constraints, we have $N_{dark\; star}\approx
5N_{bright\; star} $,  $M_{dark\; star}\approx 0.6M_{bright\; star} $ and 
 $R_{dark\; star}\approx 0.05R_{bright\; star} $
Due to the difficulty in getting a decay of DM into DE (see 
\cite{interholo}), it is
quite possible that such an object be forbiden as a star and its fate is a
Black Hole, especially in view of the lack of radiation pressure: in case
the star radiates just Dark Energy the pressure might be too small to
maintain the small star, determining its fate as a compact object. 
An object of roughly the same mass and radius 20 times smaller is $10^4$
times denser, a fact which might lead us to Black Hole formation.
This picture of an interacting theory but with supressed DE production
is a correct one to explain the observed structure.

\section{Planets}

For small objects we should have a mass $M=Nm_1$ and $R\approx
N^{1/3}a_0$, where $a_0=\frac 1{\alpha m_2}$ is the smallest quantum
size of a molecule of dark matter. The gravitational energy is
$E_g = \frac{GM^2}R\approx N^2\frac {Gm_1^2}R$, while the total
energy in the form of molecule binding energy is $E_0=N\alpha^2
m_2$. For a stable configuration the gravitational energy should not
be enough to crush matter. Therefore
\bear
E_g &=& N^2 \frac{Gm_1^2}{R(\approx N^{1/3}a_0)}=N^{5/3}\frac{Gm_1^2}{a_0}\\
&=& E_0\quad {\hbox{in order not to crush the configuration, in
the compact limit}}\nonumber 
\eear

We thus find the maximum of particles, $N_{max}$,
\bearst 
N_{dark\; max} \alpha_{DM}^2m_2\approx N_{dark\; max}^{5/3}Gm_1^2\alpha_{DM}
m_2
\eearst
thus
\bear
N_{dark\; max}&\approx& \lbrack \frac{\alpha_{DM}}{Gm_1^2} 
\rbrack ^{3/2} \approx 0.2 N_{bright\; max}\quad .
\eear

The radius and mass are given by
\bear 
R_{dark\; planet} &\approx& N_{dark\; max}^{1/3}a_0=\frac {
N_{dark\; max}^{1/3}}{\alpha_{DM}m_2}\approx 0.1 R_{bright\; planet}\\
M&\approx & N_{max} m_1\approx 0.6 M_{bright\; max}\quad .
\eear

A planet has a reasonable density as compared to a usual baryonic planet.

\section{Conclusions}

We conclude that Dark Matter forms galaxies with approximately the
same size as usual galaxies and a mass 10 times larger in case the
DM particles are roughly as heavy as protons and electrons (up to a factor
of a few) and the coupling roughly a few times the QED coupling. 
This is in accordance with observations \cite{bullet} and previous
interacting models based on interacting fluids \cite{holo,interholo}.

In such a case the substructures which can be formed are either
Black Holes or small planets. This is quite
consistent with (the lack of) further observations.

\end{document}